\documentclass[10pt, journal, compsoc]{IEEEtran}

\usepackage{float}
\usepackage{multicol}
\usepackage{cuted}
\usepackage{pifont}
\usepackage{color}
\usepackage{tcolorbox}
\usepackage{multirow}

\usepackage{cite}
\usepackage{graphicx}
\usepackage{epstopdf}

\usepackage[blocks]{authblk}

\usepackage{multirow}

\usepackage{fixltx2e}
\usepackage{subfig}

\usepackage{color}

\usepackage{url}
\usepackage{float}
\usepackage{morefloats}

\usepackage{amsmath}

\begin{document}
%

\title{Active Authentication of Keyboard Users: Performance Evaluation on 736 Subjects}


\author[1]{Md~Enamul~Karim}
\author[2]{Kiran~S.~Balagani}
\author[3]{Aaron~Elliott}
\author[4]{David~Irakiza}
\author[4]{Mike~O'Neal}
\author[5]{Vir~Phoha}
\affil[1]{mekarim@saumag.edu\\
 Southern Arkansas University, USA}
\affil[2]{kbalagan@nyit.edu\\
New York Institute of Technology, USA}
\affil[3]{aaron@aegisresearchlabs.com\\
 Aegis Research Labs LLC, USA}
\affil[4]{\{dirakiza,mike\}@latech.edu\\
 Louisiana Tech University, USA}
\affil[5]{vvphoha@syr.edu\\
 Syracuse University, USA}

\maketitle

\begin{abstract}
Keystroke timing based active authentication systems are conceptually attractive because: (i) they use the keyboard as the sensor and are not hardware-cost prohibitive, and (ii) they use the keystrokes generated from normal usage of computers as input and are not interruptive. Several experiments have been reported on the performance of keystroke based authentication using small datasets. None of them, however, study a practical active authentication system, and the feasibility of keystroke based active authentication system for large scale and continuous deployment is still not demonstrated in the literature. We investigate this issue and establish that keystroke based active authentication systems can be highly accurate and scalable. We use a real active authentication system that we developed and analyze a dataset large enough to produce statistically significant results.  We also present empirical methodologies used for characterizing various design parameters of the developed system.
\end{abstract}


%
\IEEEpeerreviewmaketitle
\section{Introduction}\label{one}
Traditional methods for user authentication, such as passwords, PINs, or fingerprint scans are designed for login-point entry and therefore cannot prevent unauthorized access to a computer after login.  A solution to this problem is in-session authentication of a user at frequent intervals, a process known as \emph{active authentication}.  Adopting conventional methods such as passwords or fingerprint scans for active authentication is impractical because they require explicit user participation and therefore will interrupt the user at every authentication event. 
For this reason, active authentication systems rely on behavioral biometrics like keystroke dynamics \cite{Killourhy} (how the user types), mouse dynamics \cite{shen2013user} (how the user uses the mouse), and/or gaze dynamics \cite{deravi2011gaze} (how the user gazes the screen). The motivation behind using behavioral biometrics is that they can be acquired from the user's natural interaction with the computer, thereby allowing the user to implicitly participate in the authentication process without getting interrupted. 

Among the various behavioral biometrics that have been considered for active authentication \cite{Gunetti, deravi2011gaze, zhao2014mobile, derawi2010unobtrusive, shen2013user}, keystroke dynamics remains one of the most studied. Though there are many papers on keystroke based active authentication (e.g., \cite{Shimshon,Gunetti,Monrose,Rahman,Bergadano2,wu}), the results reported in most of them are derived from small datasets, representing tens or at most a few hundreds of individuals (with the exception of \cite{wu}). 


Further, most studies in keystroke authentication focus their evaluations on false accept, false reject, and equal error rate metrics. These metrics, while useful for evaluating the distinctiveness of keystroke biometric (i.e., its ability to distinguish between any two users), do not fully capture the operational performance of an active authentication system. An operational active authentication system would not only seek to minimize false accept and false reject rates, but also \textit{time-to-authenticate} and \textit{time-to-unauthenticate}.  Time-to-authenticate is the period of time between two authentication events and measures how closely the system approximates `continuity' in authentication. A large time-to-authenticate translates to more time to detect an impostor, especially if the impostor hijacks the session immediately after an authentication event. Time-to-unauthenticate is the amount of time required to identify an imposter once the imposter starts typing. This metric determines how long an imposter may have access to a system guarded with keystroke based active authentication system.


\subsection{Contributions of Our Paper}

In this paper, we present a keystroke based active authentication system, which integrates multi-classifier selection, decision fusion, and threshold selection. We evaluate our system's performance on a large dataset containing keystrokes from 736 users. To our knowledge, ours is the only work that extensively tests a keystroke based active authentication system using both operational (time-to-authenticate and time-to-unauthenticate) and distinctiveness  (false accept, false reject, and half total error rate) metrics. 

We compare the performance of our system with the approaches presented in three previous studies, by Gunetti and Picardi \cite{Gunetti}, Sim and Janakiraman \cite{sim}, and Shimshon et al. \cite{Shimshon}. We show that the performance of their approaches significantly drop when tested with our dataset, reaffirming the importance of using large keystroke datasets for benchmarking. However, such datasets are not currently easily accessible to the research community. A key contribution of our work is a free text dataset from 736 users that can be used to benchmark future typing-behavior based active authentication systems. We make the dataset available to public at www.latech.edu/active-authentication.

Though thresholding is an important step in an operational active authentication system, the impact of thresholding methods on the performance of keystroke-based active authentication has not received much attention. We evaluate three thresholding methods (K-Chen, population-based HTER minimization, and user-specific HTER minimization). We show that user-specific HTER minimization outperforms the other two.  Further, we show that the performance of user-specific thresholds is stable across different population sizes (ranging between 100 and 736)  and across fixed-sized populations with non-overlapping users (population size was set to 100 users).   

 \subsection{Organization of Our Paper}

The remainder of this paper is organized as follows: Section \ref{three} presents the details of our active authentication system. Section \ref{four} presents the experiments. Section \ref{five}  presents our results. Section \ref{six} compares the performance of existing solutions to our work. We conclude in Section \ref{seven}.

\section{Details of our Approach}\label{three}

\subsection{Feature Sets and Authentication Algorithms} \label{3.1}

A keystroke timing feature is determined by the press and release times of the associated keys. It can be a uni-graph involving just one key, digraph involving two consecutive keys or multi-graph involving more than two consecutive keys. While higher graphs provide more distinguishing contexts, their availability (i.e., frequency of a multi-graph in a given text) is low which makes them less usable. In general, digraphs provide a compromise between context and availability and are popularly used for keystroke based authentication \cite{Killourhy,Rahman,Gunetti}. Although the authors in \cite{sim} claim that digraphs do not work well for free texts if their word contexts are not considered, our findings do not support that result. We considered both uni-graph and digraph features in our study. We also considered uni-graphs with adjacent key context and found them to outperform other keystroke timing features reported in the literature. Altogether, we investigated the following seven types of timing features. 

\begin{enumerate}
\renewcommand{\labelenumi}{(\roman{enumi})}
\item	Key hold latency (KH) -- the time elapsed between press and release of a given key,

\item	Inter key latency (IK) -- the time elapsed between the release of a key and press of the next key,

\item	Key press latency (KP) -- the time elapsed between the presses of two consecutive keys,

\item	Key release latency (KR) -- the time elapsed between the releases of two consecutive keys,

\item	Key hold latency with next key context (KH\textsubscript{next}) -- KH features for a given key, grouped by the next adjacent character,

\item	Key hold latency with previous key context (KH\textsubscript{prev}) -- KH features for a given key, grouped by the previous adjacent character, and

\item	Key hold latency with word context (KH\textsubscript{wc}) -- KH features for a given key, grouped by the word in which the character occurs.
\end{enumerate}



We implemented five authentication algorithms (i.e., verifiers) that have been recommended in the literature [9-11]. Two of these algorithms are distance based (SM: Scaled Manhattan \cite{Killourhy} and SE: Scaled Euclidean \cite{Killourhy}), two are matching-ratio based (A: Absolute \cite{Gunetti} and S: Similarity \cite{Rahman}) and one is order based (R: Relative \cite{Gunetti}). 

\subsection{Decision Threshold Determination}\label{3.2.2}
Once a verifier computes a score for a keystroke-sample, it has to make a decision whether that sample belongs to an genuine user or not. The standard approach is to establish a decision threshold to determine whether a score belongs to the genuine user or not, so that the system objective (e.g., authentication accuracy or EER or HTER, etc.) is optimized. While computing the thresholds we choose to optimize Half of Total Error Rate (HTER). HTER is computed by averaging False Accept Rate (FAR) and False Reject Rate (FRR). HTER approximates Equal Error Rate (EER) but is computationally less expensive and more practical \cite{Chen}. We examined three methods: 1) user-specific HTER minimization, 2) population based HTER minimization, and 3) K-Chen method, which is a hybrid method that computes user-specific thresholds using user-specific statistics and population-based parameters. Details follow.

\subsubsection{HTER Minimization Thresholding}
For each verifier-feature pair, we compute a set of scores for the training data set. Then, for computing the user-specific threshold for a given verifier-feature pair, we consider the values located between the minimum and maximum scores range for the associated user for that verifier as the candidate thresholds. Considering each of those values as thresholds we compute corresponding HTERs. The threshold that generates the lowest HTER is considered to be the final threshold. The population based HTER method computes thresholds in a similar manner, scanning all users' scores rather than one individual's score, to determine the threshold. 

\subsubsection{K-Chen Thresholding}
The K-Chen method \cite{Chen} is a user-specific thresholding method that uses user-specific statistics along with several population-based parameters. This method has been successfully used in speaker verification systems. 
The K-Chen threshold is calculated as
$T_{K-Chen} =  b(\mu'+a\sigma')+(1-b)\mu$,
where $\mu$, $\mu'$ and $\sigma'$ respectively are the mean of genuine and imposter scores and standard deviation of imposter scores. $a$ and $b$ are population based parameters, empirically computed such that the corresponding thresholds produce the best accuracies. Alternatively, these values can be set such that the resultant thresholds match with the population based HTER thresholds.


\subsection{Decision Fusion}\label{3.2.3}

Our decision to construct a fusion based system is motivated by the findings in \cite{zoo} that the biometric performance of an individual may significantly vary from use of one algorithm to another. In addition, numerous studies found that multi-classifier systems, in general, produce better decisions than a single classifier \cite{fusion3,fusion4,fusion5}. 

Our system includes five verifiers and seven types of keystroke features. Together, they form thirty-five combinations of verifier-feature pairs and produce thirty-five decisions. Fusion of these decisions adds new challenges such as (1) how to assign weight to each verifier and (2) how to deal with the decision correlations among different verifiers.  

We assign weights to verifiers' decisions using the Simultaneous Perturbation Stochastic Approximation (SPSA) algorithm  \cite{spsa}. The advantage of SPSA is that it naturally normalizes the effect of classifier correlations through the proper weighing of individual classifier decisions during decision fusion. This process makes one classifier more or less contributing than the other, and even resulting some to be totally ignored (i.e., by assigning a weight equal to 0). Further, SPSA is not affected by the dimensionality of the gradient vector as it depends on the measurement of the objective function, not on its gradient. However, because it is an approximation algorithm, it is possible that the average HTER achieved using this algorithm is not optimal. If, however, the optimal HTER is attained, changing the associated fusion threshold will not improve the average HTER further; otherwise it may. Based on this observation, once verifier decision weight assignment was complete, we scanned over false reject rate vs. false accept rate graph and re-adjusted the fusion threshold if doing so resulted in a lower average HTER.

\section{Experiments}\label{four}
\subsection{Data Set Collection}\label{4.1}

The work reported herein is based upon a large dataset collected from 736 subjects - primarily the students and faculty of Louisiana Tech University. Louisiana Tech University undertook a Institutional Review Board-approved large-scale, multi-year keystroke data collection project known as ``Typing for Ten'' and this dataset is a part of that collection.

Typing data from the 736 subjects was gathered in two phases.  In the first phase, data from 570 of the 736 subjects was collected between April 18 through May 8th 2012.  Data from the remaining 166 subjects was collected between October 15, 2012, and October 31, 2012.  The April/May and October groups are completely distinct in that no individual subject is counted as participating in both phases.

\subsection{Data Collection Protocol}\label{4.2}
\subsubsection{Two-session Data Collection}
Each of the 736 subjects in this study typed in two separate sessions: the gap between the two sessions ranged from 0 to 19 days. In each session, the participants were asked to type at least 300 characters in response to each of twelve questions presented to them. Thus the minimum number of keystrokes collected per subject per session was 3600. However, most of the subjects tried to answer the questions to the best of their ability and ended up over-typing. As a result, the average number of keystrokes collected per user per session surpassed the minimum. In Phase 1, we collected 5222 and 5085 keystrokes on average per user, in sessions 1 and 2 respectively. In Phase 2, we collected 4820 and 4704 keystrokes on average per user, in sessions 1 and 2 respectively. 


\subsubsection{Application Context}
The data were collected in a single application context - the data collection application occupied the entire screen and other functions of the host computers were locked out during data acquisition. The subjects typed their answers to the twelve questions they were asked during each session in a window-based text-editor. Once an answer was submitted, the next question popped up and the window was automatically refreshed. The order of the questions was randomized for each subject.  

\subsubsection{Acquiring Free-text Keystrokes}
The subjects typed free text, i.e., whatever they could think of in response to a given question. The questions were designed to require different levels of cognitive load, as is expected in real life scenarios. The cognitive load of individual queries is categorized according to the six levels specified in the revised Bloom's Taxonomy \cite{krathwohl2002}. The six levels of cognitive load, from the lowest to the highest, are: (1) Remembering, (2) Understanding, (3) Applying, (4) Analyzing, (5) Evaluating, and (6) Creating. For example, a participant may be asked ``What is your favorite season / time of year and what do you like about it?'' (Level 1 - Remembering); or ``Give a brief, but sufficiently detailed, plot description of your favorite book / story / movie.'' (Level 4 - Analyzing); or ``If you were to draw a picture of any type of landscape you wanted, what objects would you include in it? How would you go about drawing the landscape?'' (Level 6 - Creating).


\subsubsection{Data Anonymization}
To protect the privacy of participants, the raw data was anonymized with randomly generated IDs before being used in this project.

\subsection{Training \& Testing}\label{4.3}
\subsubsection{Training}
Our training process consisted of two steps: profile generation and parameter setting. During profile generation, 3,300 keystrokes from Session 1 were used to generate templates (profiles) of each individual by averaging their keystroke timings for those keystrokes.

During parameter setting, the rest of the keystrokes from Session 1 were used to test against the users' profiles to compute genuine scores. Additionally, Session 1 data from thirty randomly selected imposters was used per genuine user to compute impostor scores. These scores were then used to compute verification thresholds using the HTER minimization and K-Chen methods. For the K-Chen method, we used the population-based HTER thresholds for determining the population based K-Chen parameters. Once the thresholds were generated, we computed decisions for each of the scores. We then used the Simultaneous Perturbation Stochastic Approximation (SPSA) algorithm \cite{spsa} to find classifier weights for decision fusion.

\subsubsection{Testing}
Session 2 data was used to provide pristine test data completely separate from training data. All of the Session 2 genuine data per user and randomly selected imposter data from 30 other users was tested against the templates created with Session 1. Different sets of imposters were selected for each user during training and testing. Authentication decisions (genuine or imposter) were made using the thresholds computed during the training phase.  Finally, decision fusion was performed using the classifier weights and the fusion threshold computed during training. 


\subsubsection{Test setup for computing time-to-unauthenticate} 
\label{time-to-authen}
For computing time-to-unauthenticate, we simulated transition between genuine and imposter samples by interleaving a genuine sample among 30 imposter samples collected from 30 different randomly selected imposters. For each user, we created the following sample sequence: 
\begin{equation*}
G_{A} - I_{A,1} - G_{A} - I_{A,2} - \cdots - G_{A} - I_{A, 30}
\end{equation*}
where $G_{A}$ refers to the keystroke samples from a genuine user $A$ and $I_{A,i}$ refers to the keystroke samples from impostor $i$ ($i \neq A$). Time between the last keystroke of an individual and the first keystroke of the next individual was ignored since in reality two individuals did not type continuously one after another. As explained in Section \ref{4.10ok}, the first authentication decision is made after the first sample (550 keystrokes) is presented and the subsequent decisions are made at $55^{th}$ keystroke thereafter. 


\section{Results}\label{five}

\begin{table*}[htpb]
\centering
\caption{Average HTERs achieved with 35 different verifier-feature combinations and fusion.}
\begin{tabular}{|c|c|c|c|c|c|c|c|}
\hline
& \textbf{IK} & \textbf{KH} & \textbf{KPL} & \textbf{KRL} & \textbf{KH\_Next} & \textbf{KH\_Prev} & \textbf{KH\_WC} \\\hline 
\textbf{Absolute} & 0.220249 &  0.175669  & 0.178495 & 0.145255 & 0.08576 & 0.165354 & 0.137037 \\\hline
\textbf{Relative} & 0.110346 &  0.141892  & 0.140179 & 0.11691 & 0.097149 & 0.162492 & 0.188772 \\\hline
\textbf{Scaled Euclidean} & 0.232715 &  0.07778  & 0.266054 & 0.235706 & \textbf{0.070852} & 0.095591 & 0.133891 \\\hline
\textbf{Scaled Manhattan} & 0.166021 &  0.086954  & 0.173984 & 0.164906 & 0.095727 & 0.099156 & 0.1318754\\\hline
\textbf{Similarity} & 0.319568 &  0.239606  & 0.270084 & 0.197835 & 0.118748 & 0.239804 & 0.153775 \\\hline \hline
\textbf{Fusion with User-Specific Thresholding} & \multicolumn{7}{|c|} {\textbf{0.010607411}  (FAR: 0.007921923; FRR: 0.013292899)}\\ \hline
\textbf{Fusion with K-Chen Thresholding} & \multicolumn{7}{|c|} {0.014581887 (FAR: 0.015217012; FRR: 0.014581887)}  \\ \hline
\textbf{Fusion with Population-based Thresholding} & \multicolumn{7}{|c|} {0.023132105 (FAR: 0.01897904; FRR: 0.023132105)} \\ \hline
\end{tabular}
\label{table_ind}
\end{table*}


Table \ref{table_ind} shows the average of user-specific HTERs for each of the thirty-five verifier-feature pairs we tested.  The lowest average HTER of any individual classifier is about .07 -- for ``$Scaled Euclidean-KH_{next}$'' verifier-feature pair. Table \ref{table_ind} also presents the averages of user-specific error rates after SPSA-based fusion was performed and decision thresholds are adjusted. It shows that the user-specific HTER threshold based scheme, clearly, is the winner.

In addition to achieving 0.01 HTER, user-specific HTER thresholding yielded 0.008 FAR and 0.013 FRR.  This means it is less likely to err towards reporting an imposter as genuine than to report that a genuine user is an imposter.  In other words, it is more sensitive to reporting actual imposters, at the cost of being a bit more likely to raise false alarms.

\begin{figure}[ht]
\begin{minipage}{\columnwidth}
\centering
\includegraphics[scale=0.75]{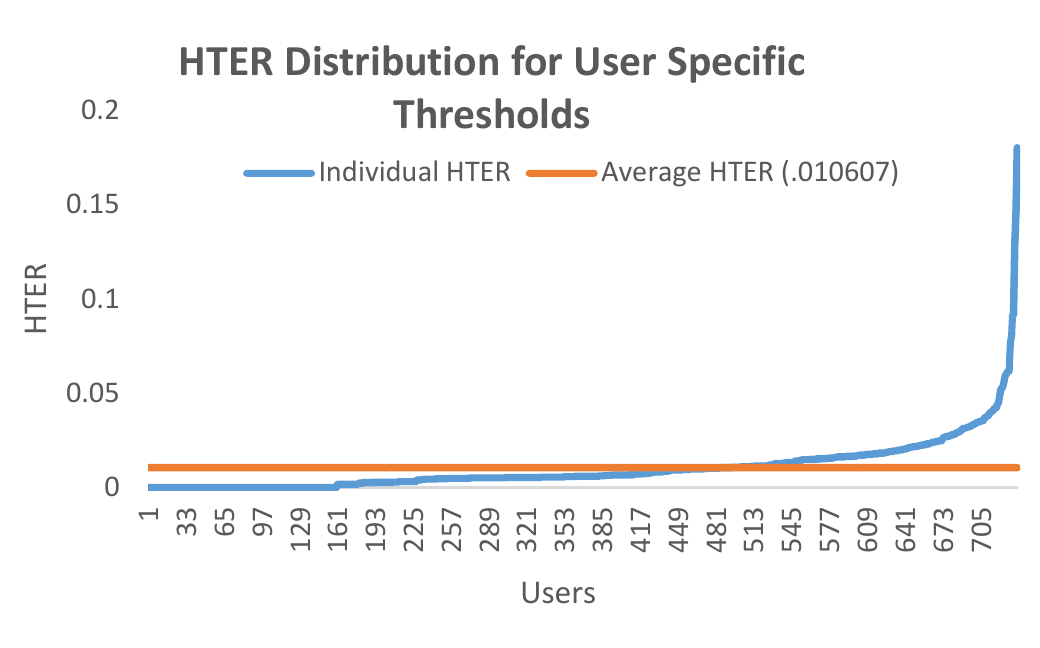}\\
(a)
\end{minipage}
\begin{minipage}{\columnwidth}
\centering
\includegraphics[scale=.75]{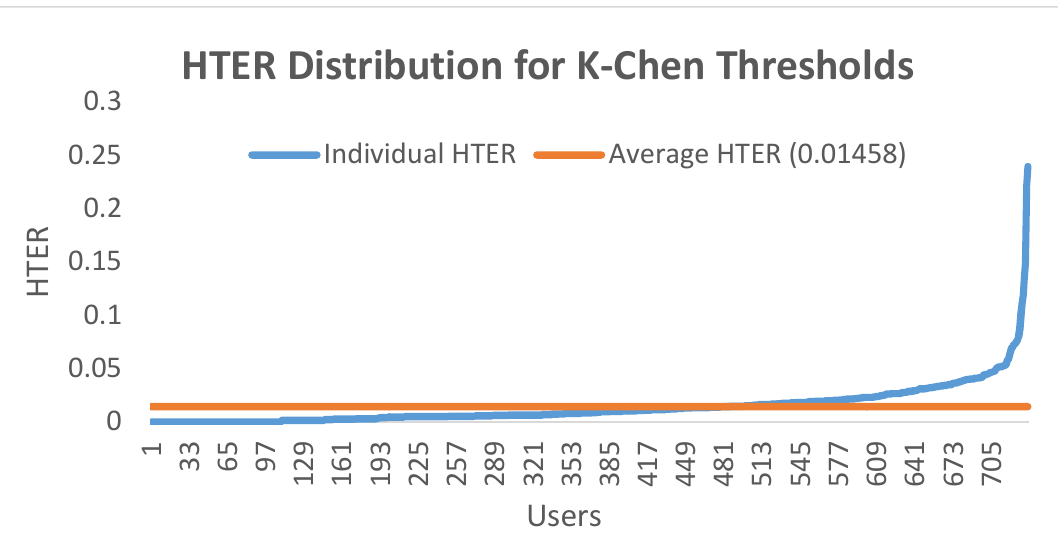}\\
(b)
\end{minipage}
\begin{minipage}{\columnwidth}
\centering
\includegraphics[scale=.75]{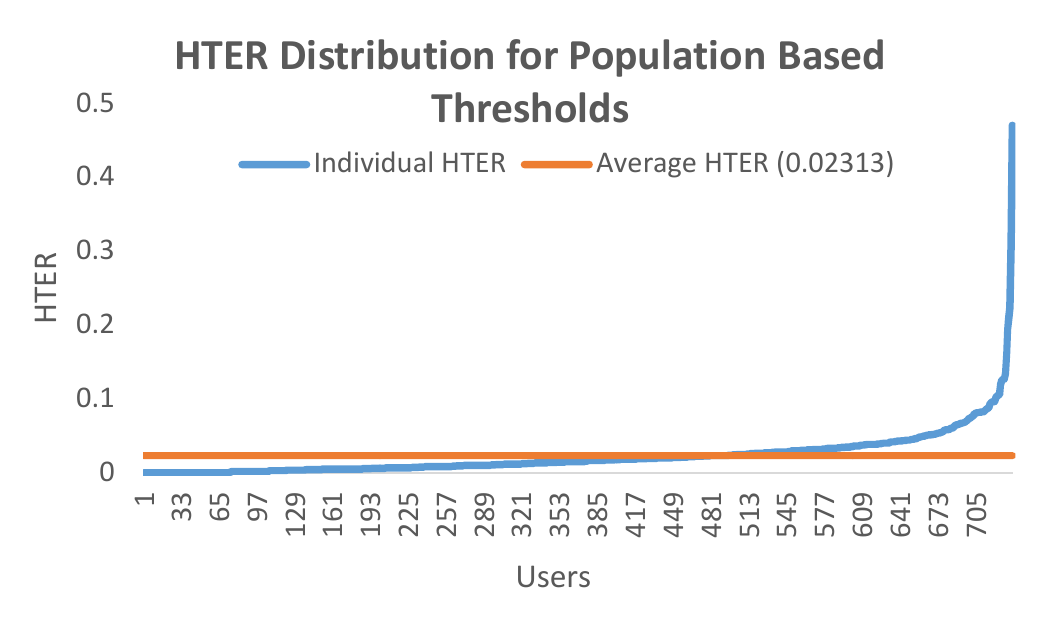}\\
(c)
\end{minipage}
\caption{HTER distributions for three thresholding methods.}
\label{ana1}
\end{figure}

Figure \ref{ana1}  explains the overall error characteristics of the three threshold schemes we studied. Figure \ref{ana1} shows that for each of the schemes most of the users perform better than the average HTER while a few users have very high HTERs. Although the list of the worst 20\% users significantly overlapped in all schemes, their relative performance varied between schemes. For instance, the worst HTER scores under the population-based HTER threshold were generated by subject 1070 (HTER of 0.47 which is essentially equivalent to random guesses) and subject 1974 (HTER of 0.31). These same two individuals were also the worst performing under the K-Chen method, though their HTERs improved substantially -- 0.22 for subject 1070 and 0.13 for subject 1974. Interestingly, the worst two HTER scores under the user-specific HTER method belonged to different subjects: subject 485 with an HTER of 0.18 and subject 1515 with an HTER of 0.1469. Thus, not only does user specific HTER outperform the other two methods in the average case, it substantially outperforms them even on the small number of poorly performing subjects.

Though the user-specific HTER threshold scheme yielded the best HTER in 572 out of 736 cases, in 97 cases K-Chen method outperformed the user-specific method and in 58 cases the population based HTER method outperformed the user-specific method. Additionally, in 9 cases both K-Chen and population based HTER methods outperformed the user-specific method. Given these results, an interesting proposal for future investigation would be to vary the threshold selection method on a subject by subject basis to improve the overall system accuracy.


While inspecting the typed texts of the users with the worst HTERs, we did not notice any anomalies except for the subject 1070. Subject 1070, most of the time, instead of typing meaningful answers to the the given questions, typed random characters, such as the following: `` GH G G G G HG HG HG G UH UY UYUY UYU YUY UY UY Y Y Y Y...''

With population based HTER thresholding, the FRR, FAR and HTER for subject 1070 respectively were, 0.94, 0, and 0.470760234 respectively. Which means, almost all of the tested samples were identified as imposter samples. This happened because the mechanical nature of the typed text most likely deviated from the population characteristics. When user-specific HTER thresholds were used, the profile created using ``mechanical'' timings became closer to the timings of the test samples that were also characterized by these ``mechanical'' timings. 



\subsubsection{A Closer Look at the Worst HTER Quintile}\label{4.8}
Multiple recent studies \cite{zoo} \cite{Yager} demonstrate that the average performance of a given biometric system is disproportionately affected by a small segment of users. We find the same to be true for keystroke based biometrics. In our study, with user-specific HTER thresholding,  21.74\% of the subjects generated no error, 66.4\% of the subjects performed better than the average (HTER .01). The remaining (small) percentage of the subjects had high HTERs, which primarily drove average. 

\begin{figure*}[htpb]
\centering

\subfloat[Distribution of HTER for the test set]{%
 \label{figurex}
  \includegraphics[scale=0.75]{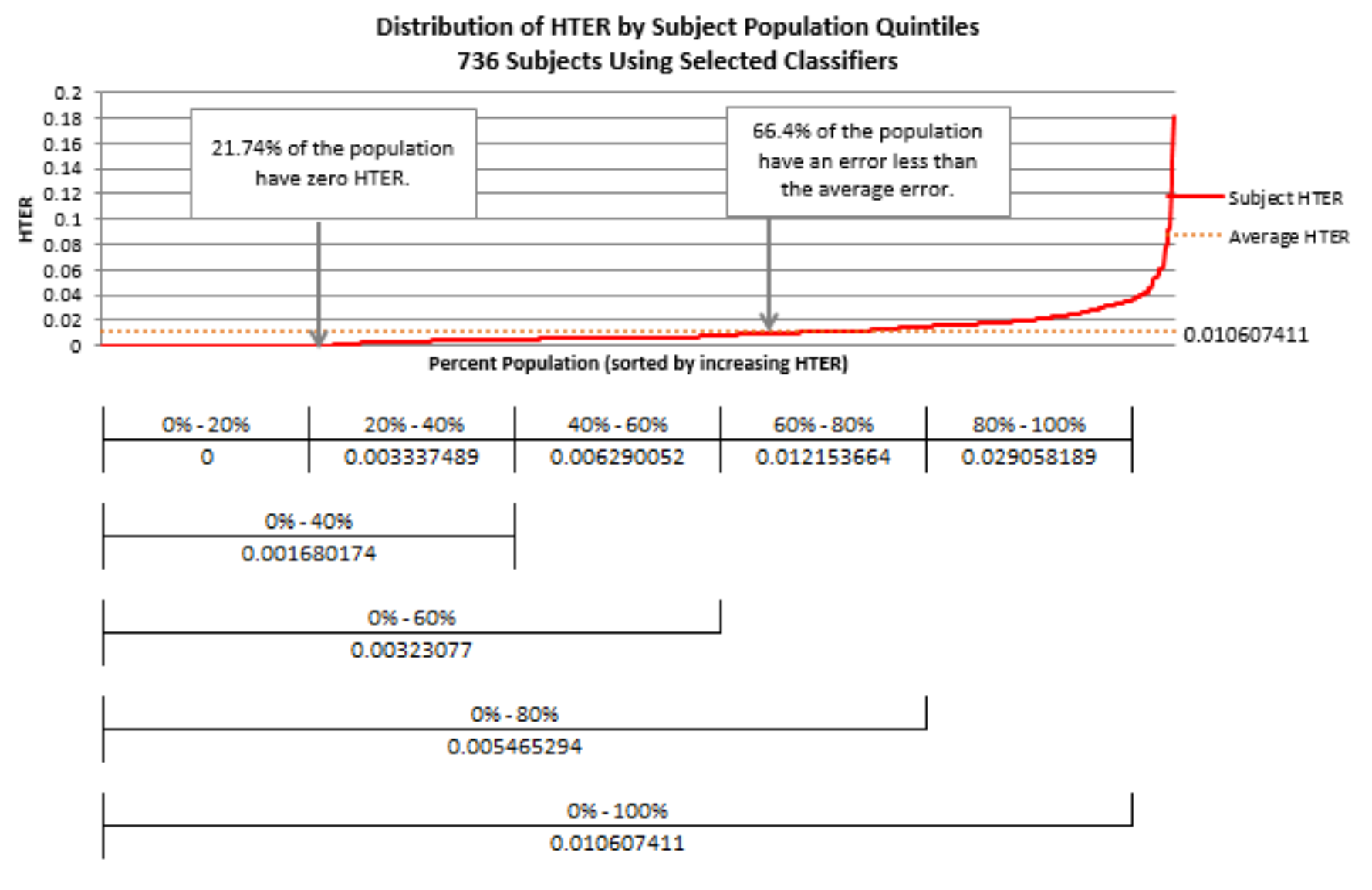}%
}

\subfloat[Distribution of HTER for the worst 20\% performers]{%
  \label{figurey}
  \includegraphics[scale=0.75]{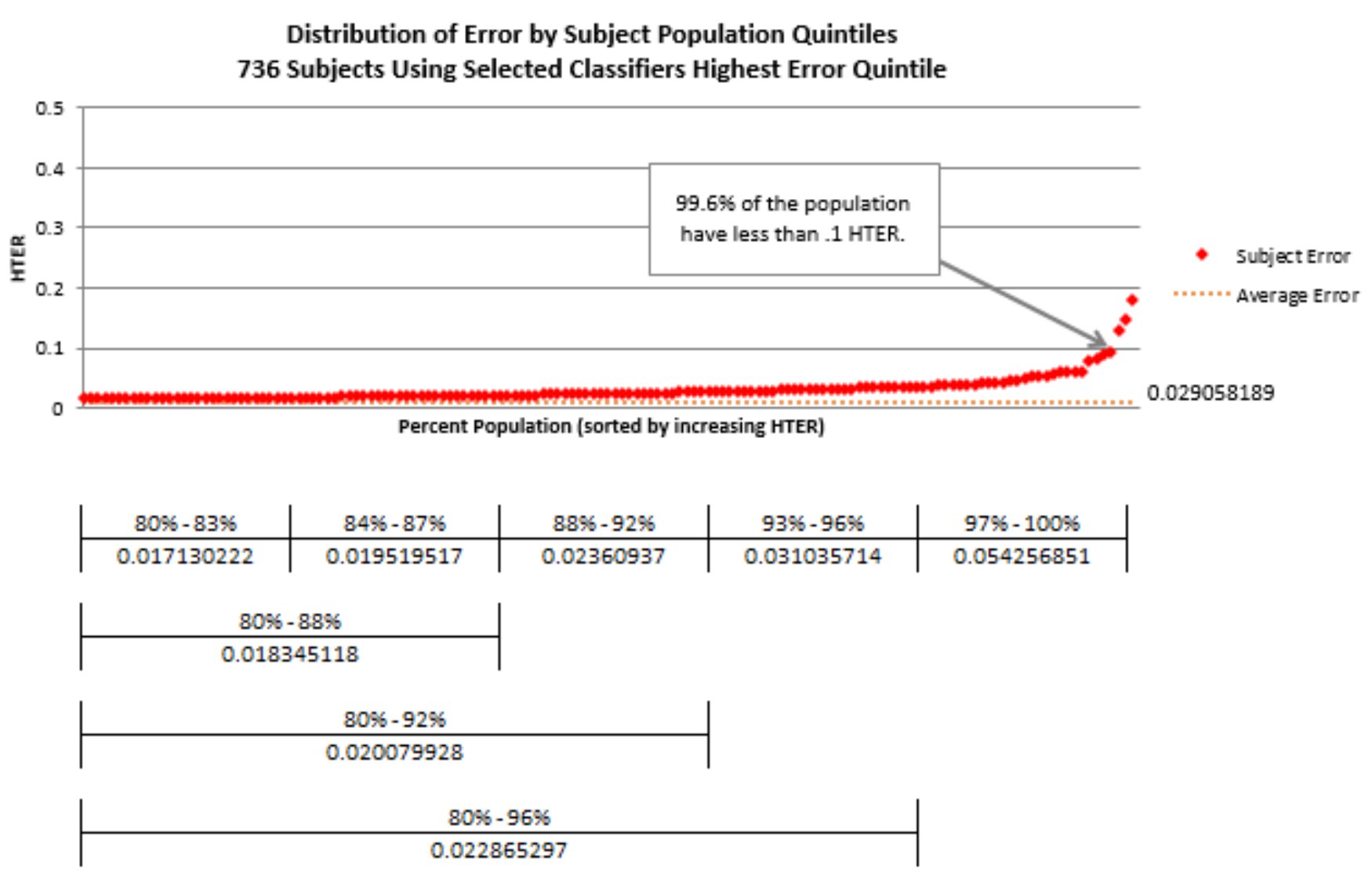}%
}

\caption{Distribution of HTERs}

\end{figure*}


Figure \ref{figurex} shows the distribution of HTERs by population quintiles, when we used user-specific HTER thresholding.
A closer look to the worst quintile of the performers in Figure \ref{figurey} reveals that 99.6\% of the population have an overall HTER of less than 0.1. In fact, only three out of 736 users had an HTER greater than 0.1 and only one had HTER greater than .15.

\subsection{Stability of User-specific HTER Minimization Thresholds}\label{4.7}
We further tested the stability of user-specific HTER minimization thresholding across different sizes of user populations. Table \ref{table11} shows how the overall error rates are impacted when the user-specific HTER threshold scheme and fusion was used on (a) non-overlapping  (i.e., users in each group do not overlap) populations of identical size and (b) populations of different sizes. The table reports the standard deviation of HTERs in the second and fourth columns. The low standard deviation in Column 2 indicates that the HTERs are stable across groups of 100 users. The low standard deviation in Column 4 indicates that the HTERs are stable across different population sizes. The table also reports the net deviations of HTERs. To calculate net deviation, we used standard deviation formula with the HTER for 736 users (i.e., 0.010607412) replacing the column-wise mean of HTERs. The low net deviations indicate that despite the different user populations and the population sizes, the HTERs remain consistently close to the HTER obtained with 736 users. 

\begin{table}[ht]
\caption{System performance for various subsets of population}
\begin{tabular}{|c|c|c|c|}
\hline
\textbf{Non-overlapping} & \textbf{HTER} & \textbf{Cumulative \#} & \textbf{Cumulative}\\
\textbf{\# of Participants}&&\textbf{of Participants}& \textbf{HTER}\\\hline
100&0.010700883&100&0.010700883\\\hline
100&0.010413961&200&0.010557422\\\hline
100&0.010511376&300&0.010542073\\\hline
100&0.010844217&400&0.010617609\\\hline
100&0.010485292&500&0.010591146\\\hline
100&0.010700003&600&0.010609289\\\hline
100&0.01053229&700&0.010598289\\\hline
36&0.010784816&736&0.010607412\\\hline 
\multicolumn{2}{|r|}{$\mu$: 0.010621605}&\multicolumn{2}{|r|}{ }\\
\multicolumn{2}{|r|}{$\sigma$: 0.000156045}&\multicolumn{2}{|r|}{}\\\cline{1-2}
\multicolumn{2}{|r|}{Net Deviation: 0.000146655}&\multicolumn{2}{|r|}{Net Deviation: 0.000047846}\\\hline

\end{tabular}
\label{table11}
\end{table}




\subsection{Separation between Training and Testing Sessions}\label{4.9}


\begin{figure}[ht]
\centering
\includegraphics[width=\columnwidth]{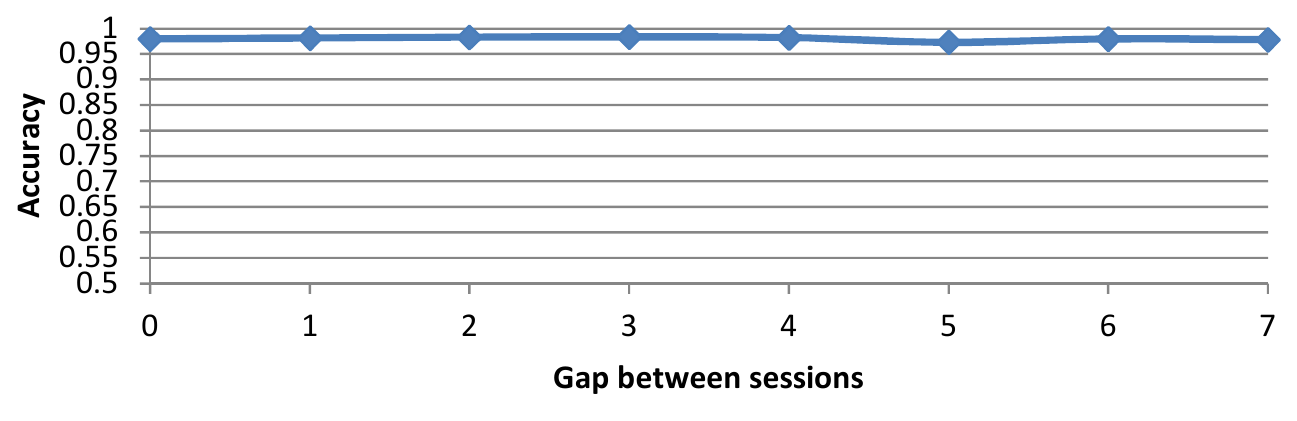}
\caption{Accuracy versus gap (number of days) between training and testing sessions}
\label{figure6}
\end{figure}

Each of the subjects included in our study typed in two sessions and the gap between the two sessions ranged from 0 to 19 days. The vast majority of the subjects (all but 47 of the 736 users) returned within 7 days to type the second session. When we grouped users by 1 through 7 days of separation between training and testing sessions, we observed that the average authentication accuracies across the user groups remained almost same, indicating the keystroke signatures were consistent over multiple days. There was insufficient data to look beyond the one week horizon.

Figure \ref{figure6} shows the average accuracies for users grouped by 1 through 7 days of separation between sessions.  As shown in Figure \ref{figure6}, the keystroke signatures of our test subjects were consistent over multiple days, with no apparent drop in system accuracy over the course of a week.  There was insufficient data to look beyond the one week horizon.

\subsection{Time-to-authenticate and Time-to-unauthenticate}\label{4.10ok}

We observed that authentication performance of our system saturates around  550 keystrokes. However, it takes on average 200 seconds for an average typist to type that many keys. To avoid this delay without compromising accuracy, we use a sliding window of 550 keystrokes. During each authentication decision, we move the window by 1/10th of its size, i.e., we discard 55 keys from the window and add 55 keys to it in a FIFO manner. Thus after the first 200 seconds, time-to-authenticate remains about 20 seconds while a user is typing and authentication is always made with 550 keystrokes.

\begin{figure}[ht]
\centering
\includegraphics[width=\columnwidth]{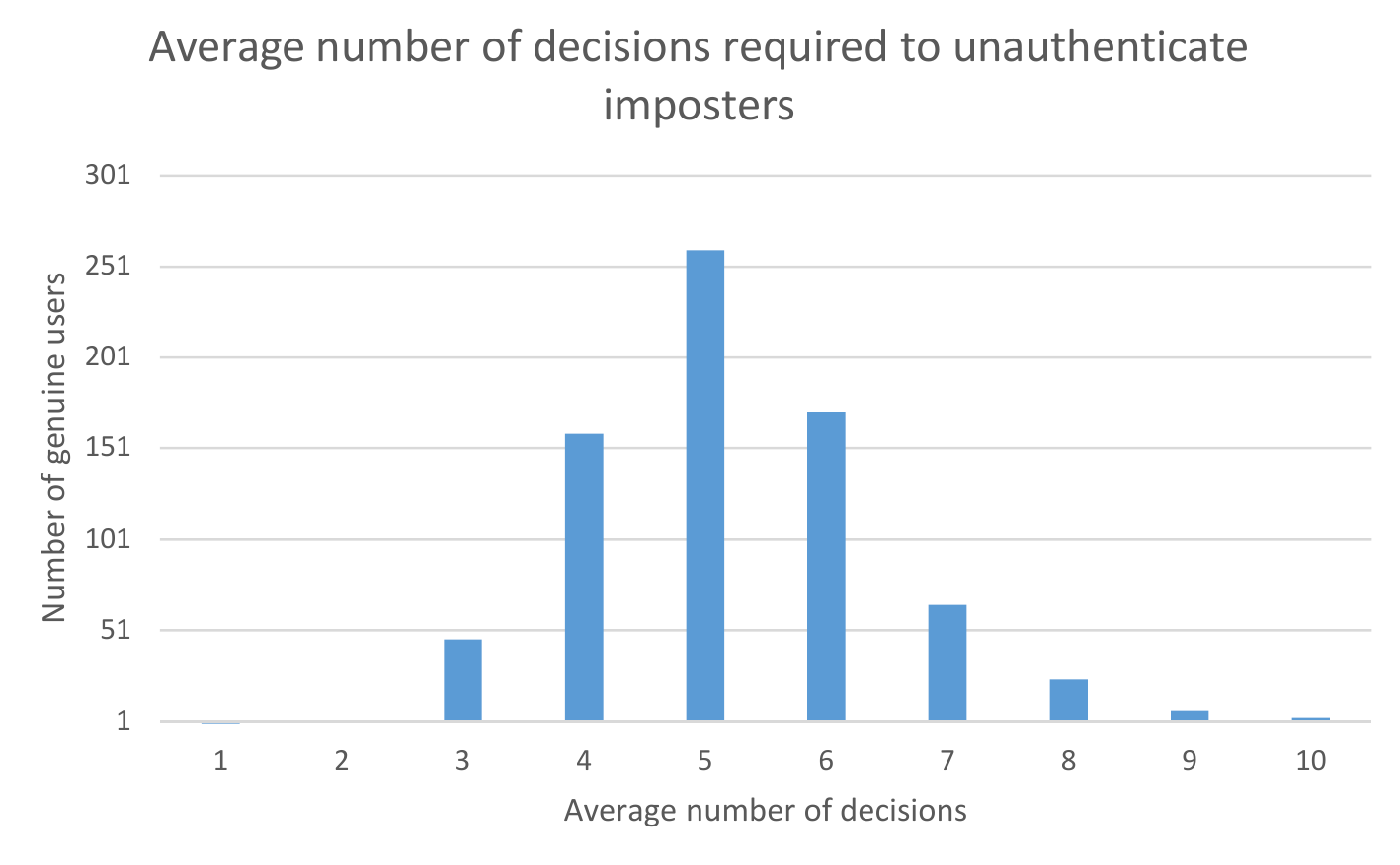}
\caption{Histogram for average number of decisions required to unauthenticate impostors}
\label{deci}
\end{figure}

In order to determine time-to-unauthenticate, we follow the test protocol discussed in Section \ref{time-to-authen} and used a sliding window of  550 keystrokes. We identified 92.25\% of the imposters within first 7 decisions i.e., 385 keystrokes. In Figure \ref{deci}, we show a histogram of the distribution of average decisions required to unauthenticate an individual. 

\section{Comparison with Existing Approaches}\label{six}

Gunetti and Picardi \cite{Gunetti} developed two algorithms for free text based authentication that we included in this study. Using Absolute and Relative algorithms on 205 individuals they obtained a False Rejection Rate of less than 5\% and a False Accept Rate of less than 0.005\%. However, using our dataset, the accuracy of their technique drops to 93.34\% with an FRR of 6.61\% and FAR of 6.59\%.

Sim and Janakiraman \cite{sim} observed that if the digraphs are fine-tuned with their word contexts, then, the accuracies improve. Consideration of word contexts means that the ``in'' in ``Bin'' and ``in'' in ``China'' are two different features. While there are advantages to taking this approach, it exacerbates the issue of feature availability (e.g., the character combination ``in'' generally appears much more frequently than do the words ``Bin'' or ``China''). Replicating their method the best accuracy we were able to attain was 86.29\%.  This disappointing result was due to a paucity of feature availability.

Shimshon et al. \cite{Shimshon} addressed the ``time-to-authenticate'' issue in keystroke based active authentication. They proposed the use of a group for digraphs with similar timings as substitutions for missing features. That way, when a digraph from a group is available, the timing for the other digraphs (that are not typed) in that group can be estimated resolving the feature availability issue and thus shortening the time-to-authenticate period. However, doing so tends to degrade the quality of the features. Although they claim that their method performed better than existing methods, their evaluation was performed based on only 21 legitimate users. In our study we were able to attain an accuracy of only 89.4\% using their method.


\section{Conclusion and Future Work}\label{seven}

We presented a keystroke based active authentication system and validated its performance on a large dataset of 736 subjects. We also release this dataset to the research community to benchmark future active authentication systems. Our experimental evaluations show that, with multi classifier fusion and user-specific thresholding, the authentication system was able to achieve as low as 0.0106 HTER with 20 seconds time-to-authenticate (after the first 200 seconds) and requires an average of five attempts to unauthenticate. Our future work will focus on the following directions: (1) \noindent\emph{Building Resistance to Spoof Attacks:} If someone's typing data were captured by a key-logger it might be possible for an attacker to create a program that simulated the atomic keystroke behavior of that user (see \cite{Rahman}).  We are exploring the use of pause and revision behavior, and demographic and cognitive traits (extracted from the keystroke input) to detect such attacks; and (2) \noindent\emph{Time Drift:} Although we inspected the impact of a week's delay between profile generation and actual use of the system for authentication - finding no performance degradation over this limited time period, we next intend to determine the stability of our approach over extended time periods (months to years).

\section*{Acknowledgment}
This work is supported by DARPA Active Authentication Grant Number FA8750-13-2-0274. The authors would like to thank Azriel Richardson, John Hawkins, Daniel Adams, Christian Dean, Andrew Duryea, Nick Henry, Sean Manteris and Anna Whitaker
for their assistance in performing various tests for this paper.





\bibliographystyle{IEEEtran}

\sloppy
\bibliography{AA}
\end{document}